# DFT calculation for adatom adsorption on graphene sheet as a prototype of carbon nano tube functionalization

**Akira Ishii, Masana Yamamoto, Hiroki Asano and Katsutoshi Fujiwara**

Department of Applied Mathematics and Physics, Tottori University

Koyama, Tottori 680-8552, Japan

ishii@damp.tottori-u.ac.jp

**Abstract**. DFT calculation of various atomic species on graphene sheet is investigated as prototypes for formation of nano-structures on carbon nanotube (CNT) wall. We investigate computationally adsorption energies and adsorption sites on graphene sheet for a lot of atomic species including transition metals, noble metals, nitrogen and oxygen, using the DFT calculation as a prototype for CNT. The suitable atomic species can be chosen as each application from those results. The calculated results show us that Mo and Ru are bounded strongly on graphene sheet with large diffusion barrier energy. On the other hand, some atomic species has large binding energies with small diffusion barrier energies

## 1. Introduction

The carbon nanotube(CNT) is an unique material for nanotechnology. The interesting point of CNT for nanotechnology is that we can use both the inner side and the outer side of the wall of the tube. The interaction between nanostructures on the inner side and the outer side can be expected. For such purpose, adsorption on CNT wall should be investigated in detail.

The production of the CNT has become possible by the growth techniques such the carbon arc discharge method [1], the catalytic chemical vapor deposition (CVD) method [2] and PLD [3]. Thus, it will be a good timing to consider how to construct nano structure on CNT.

As a target for DFT calculation, very heavy computational work is required for realistic size CNT. Thus, as a first step to investigate adsorption of extra atoms or molecules on CNT, we calculate adatom on graphene sheet as a prototype for CNT. We can expect that qualitative feature of adsorption is same between graphene and CNT. Thus we expect to be able to apply the result for graphene for selection of target adsorbate atomic species or molecular species for nano-structure on CNT.

Thus, in this paper, we investigated adsorption energies, stable sites and diffusion barrier energies for a lot of atomic species in order to find suitable atomic species for nano structure formation.
.

## 2. Method

First-principles total energy calculations have been performed using the program packages, "VASP" and "STATE" (Simulation Tool for Atom Technology) developed at National Institute of Advanced Industrial Science and Technology (AIST). The calculations are based on the DFT using the pseudopotential plane-wave method with the LDA and GGA[4,5] for the exchange-correlation energy. We use the ultrasoft pseudopotentials and the energy cutoff is set to 42Ry. The primitive first Brillouin Zone was sampled with 4k-points.

In the case of adsorption on graphene sheet, the atomic species marked in tables 1 and 2. The unit cell for graphensheet have been adopted 3×3 structure. The distance between graphensheets is about 13.4Å and the distance between adatoms is about 7.2Å. For the calculation of adatom at certain sites, the position coordinate of adatom parallel to the surface is fixed and the coordinate normal to the surface is fully relaxed. One atom of the edge of 3×3 structure of the graphene sheet is fixed during the relaxation of other carbon atoms of the sheet. The calculations are done for adatom at three site having symmetry, H6, B and L shown in figure 1.

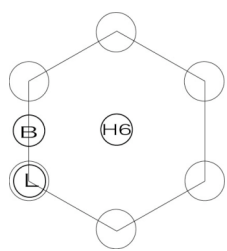

Figure 1
The three adsorption sites of graphene sheet, L, B and H6.

## 3. Results

We calculate the adsorption energy for adsorbate atoms for some atomic species shown in tables 1 and 2. The relative and absolute adsorption energies for each site are shown in table 3. From the energy difference between the adsorption energies for each adsorption sites, we can estimate the minimum limit of the migration energy of the adatom.

In the calculation, we found that the adsorption types can be divided into two types: the type of fixed adsorption site and the type of no fixed adsortion site. The type of no fixed adsorption site has the migration barrier energy less than 0.4eV. The 0.4eV corresponds roughly to the threshold energy of the atomic migration at room temperature. Thus, the atomic species in table 2 can migrate even at room temperature.

adsorption site

| | 1 | 2 | 3 | 4 | 5 | 6 | 7 | 8 | 9 | 10 | 11 | 12 | 13 | 14 | 15 | 16 | 17 | 18 |
|---|---|---|---|---|---|---|---|---|---|---|---|---|---|---|---|---|---|---|
| | H | | | | | | | | | | | | | | | | | He |
| | Li | Be | | | | | | | | | | | B | **C** | **N** | **O** | F | Ne |
| | Na | Mg | | | | | | | | | | | Al | Si | P | S | Cl | Ar |
| | K | Ca | Sc | Ti | **V** | **Cr** | **Mn** | **Fe** | **Co** | Ni | Cu | Zn | Ga | Ge | As | Se | Br | Kr |
| H6 | Rb | Sr | Y | Zr | Nb | **Mo** | Tc | **Ru** | Rh | Pd | Ag | Cd | In | Sn | Sb | Te | I | Xe |
| **B** | Cs | Ba | La | Hf | Ta | W | Re | Os | Ir | Pt | Au | Hg | Tl | Pb | Bi | Po | At | Rn |

Table 1
Atomic species having large migration energy are shown as the hatched and the void letters.

| 1 | 2 | 3 | 4 | 5 | 6 | 7 | 8 | 9 | 10 | 11 | 12 | 13 | 14 | 15 | 16 | 17 | 18 |
|---|---|---|---|---|---|---|---|---|---|---|---|---|---|---|---|---|---|
| *H* | | | | | | | | | | | | | | | | | He |
| *Li* | Be | | | | | | | | | | | B | C | N | O | F | Ne |
| *Na* | Mg | | | | | | | | | | | *Al* | *Si* | P | S | Cl | Ar |
| *K* | *Ca* | *Sc* | *Ti* | V | Cr | Mn | Fe | Co | *Ni* | *Cu* | *Zn* | *Ga* | *Ge* | *As* | *Se* | *Br* | Kr |
| Rb | Sr | Y | Zr | Nb | Mo | Tc | Ru | Rh | *Pd* | *Ag* | Cd | *In* | Sn | Sb | Te | I | Xe |
| Cs | Ba | La | Hf | Ta | W | Re | Os | Ir | *Pt* | *Au* | Hg | Tl | Pb | Bi | Po | At | Rn |

Table 2
Atomic species having migration barrier energy less than 0.4eV.

In table 1, The stable site for atomic species of transition metals having large migration barrier energy is V, Cr, Mo, Mn, Fe, Ru and Co is H6. For carbon, nitrogen and oxygen, the stable site is B.

The atomic species indicated in table 2 have very low migration energy so that the adsorption site is not stable at room temerature. As a typical example for the low migration barrier energy, we calculate the contour map of the potential surface for Pt adatom on graphene sheet as figure 2. Since the variation of the potential between the B and the L is monotonic, the migration barrier energy for Pt adatom is 0.18eV.

The displacements of carbon atoms of the graphene sheet near the adatom are within 0.3A for the direction normal to the sheet. Thus, for graphen, the strain due to the adsorption is not large.

|  | atom | H6 | B | L | Ebond |  |
|---|---|---|---|---|---|---|
| I | H | 0.004 | 0.002 | 0.000 | 0.48 | GGA |
|  | Li | 0.00 | 0.33 | 0.35 | 2.02 | GGA |
|  | Na | 0.00 | 0.11 | 0.13 | 1.07 | GGA |
| Alkali metal | K | 0.00 | 0.09 | 0.10 | 0.73 | LDA |
| Alkaline earth metal | Ca | 0.00 | 0.17 | 0.18 | 0.57 | LDA |
|  | Sc | 0.00 | 0.36 | 0.45 | 2.00 | LDA |
|  | Ti | 0.00 | 0.28 | 0.09 | 2.97 | GGA |
|  | V | 0.00 | 1.08 | 1.22 | 3.65 | LDA |
|  | Cr | 0.00 | 1.42 | 1.49 | 3.73 | LDA |
|  | Mo | 0.00 | 1.41 | 1.60 | 5.51 | GGA |
|  | Mn | 0.00 | 1.36 | 1.35 | 3.61 | LDA |
|  | Fe | 0.00 | 1.23 | 1.09 | 3.80 | LDA |
| Transition metal | Ru | 0.00 | 0.75 | 0.72 | 4.03 | GGA |
|  | Co | 0.00 | 0.82 | 0.80 | 3.37 | LDA |
|  | Ni | 0.00 | 0.05 | 0.16 | 2.09 | GGA |
|  | Pd | 0.20 | 0.00 | 0.08 | 1.15 | GGA |
|  | Pt | 0.80 | 0.00 | 0.18 | 2.35 | GGA |
|  | Cu | 0.25 | 0.00 | 0.01 | 0.54 | GGA |
|  | Ag | 0.15 | 0.00 | 0.01 | 0.18 | GGA |
|  | Au | 0.11 | 0.04 | 0.00 | 0.43 | GGA |
| II | Zn | 0.00 | 0.08 | 0.08 | 0.19 | LDA |
|  | Al | 0.00 | 0.18 | 0.17 | 1.84 | GGA |
|  | Ga | 0.00 | 0.17 | 0.18 | 1.62 | GGA |
| III | In | 0.00 | 0.19 | 0.20 | 1.81 | GGA |
|  | C | 1.66 | 0.00 | 0.61 | 3.23 | GGA |
|  | Si | 0.24 | 0.00 | 0.16 | 1.69 | GGA |
| IV | Ge | 0.26 | 0.00 | 0.12 | 1.45 | GGA |
|  | N | 3.33 | 0.00 | 0.86 | 4.41 | GGA |
| V | As | 1.31 | 0.00 | 0.20 | 1.51 | GGA |
|  | O | 2.79 | 0.00 | 0.58 | 4.42 | GGA |
| VI | Se | 0.99 | 0.00 | 0.23 | 1.44 | LDA |
| VII | Br | 0.05 | 0.01 | 0.00 | 0.28 | LDA |

Table 3 Calculated adsorption energies. Calculations are performed with LDA or GGA as indicated in the columm.

## 4. Disucssion

Though carbon nano tube has a lot of variation of radius and chiral vectors, the character of the binding of adatom is expected to be roughly similar between CNT and graphene. For Oxygen atom, the detailed first-principles calculation has been reported by S.Dag et. al.[6]. According to their calculation, the adsorption sites for Oxygen on CNT are B site which is same as our calculation for graphene. The calculated binding energies are also roughly similar to our binding energy for Oxygen on graphene. Thus, we can expect that our calculated results for graphene can be used as a reference to consider the candidate atomic specie for adsorbate on CNT.

For many transition metal species, we found that they have rather strong binding energy, but diffusion barrier energies are small, because the difference of the value at L, B and H6 sites are not. According to figure 2 for Pt, we find that the diffusion barrier energy of Pt adatom is determined from the energy difference between L and B. Thus, such atomic species, Ru, Ti, Ni, Pd, and Pt can be used as functional material to coat uniformly both inside and outside of the wall of SWNT.

We also found the strong adsorption energy and the large migration barrier energy for V, Cr, Mn, Fe, Co, Mo and Ru; the transition metals having half-filled d-band. Such atomic species can be used as coating of the outside wall of the CNT or forming nano structures on CNT. Because of the strong binding energy especially for Mo, we can expect to sit Mo atoms into CNT by breaking the wall of the CNT.

For Nitrogen and Oxygen, their diffusion barrier energy is expected to be large. Therefore, nitrogen and oxygen inside of the SWNT cannot diffuse toward the axial direction of the tube. For outside of the tube, nitrogen and oxygen adatom can adsorb strongly without diffusion.

For noble metals, as we can see in table 3, their binding energy is very small and the diffusion barrier energies are also expected to be very small. Thus, Cu, Ag and Au can be expected to flow easily in SWNT, if the radius of SWNT is large enough compared to the bond length of the noble metal atoms with the wall of SWNT. Using such property, we can use SWNT as pipe to flow noble metal atoms

## 5. Conclusion

We calculate the adsorption of extra adatom on graphene sheet as a prototype of the adsorption of extra adatom on SWNT using the first-principles calculation. The result shows us that atomic species in same group has same stable site on graphene. The transition metals of V, Cr, Mn, Fe, Co, Mo and Ru has strong adsorption energy and strong migration barrier energy. Ru, Ti, Ni, Pd, Pt, Si and Ge are suitable as material for coating of SWNT. The noble metals can flow inside of SWNT. Nitrogen and Oxygen have large diffusion barrier energy on SWNT.

**Acknowledgements**
The authors are grateful for Prof. Y. Morikawa of Osaka University for his permission to use his ab initio program codes STATE